\documentclass[11pt,a4paper]{article}

\usepackage[utf8]{inputenc}
\usepackage{amsmath}
\usepackage{amsfonts}
\usepackage{amssymb}
\usepackage{graphicx}
\usepackage{hyperref}
\usepackage{geometry}
\usepackage{authblk}
\usepackage{cite}
\usepackage{booktabs}

\title{\textbf{Be Water: An Evolutionary Proof for Trend-Following}}

\author{
  YIJIA CHEN \\
  Independent Researcher \\
  \texttt{chenyijia202@foxmail.com}
}
\date{\today}

\begin{document}

\maketitle

\begin{abstract}
The proliferation of diverse, high-leverage trading instruments in modern financial markets presents a complex, "noisy" environment, leading to a critical question: which trading strategies are evolutionarily viable? To investigate this, we construct a large-scale agent-based model, "MAS-Utopia," comprising 10,000 agents with five distinct archetypes. This society is immersed in five years of high-frequency data under a counterfactual baseline: zero transaction friction and a robust Unconditional Basic Income (UBI) safety net. The simulation reveals a powerful evolutionary convergence. Strategies that attempt to fight the market's current—namely Mean-Reversion ("buy-the-dip")—prove structurally fragile. In contrast, the Trend-Following archetype, which adapts to the market's flow, emerges as the dominant phenotype. Translating this finding, we architect an LLM-driven system that emulates this successful logic. Our findings offer profound implications, echoing the ancient wisdom of "Be Water": for \textbf{investors}, it demonstrates that survival is achieved not by rigid opposition, but by disciplined alignment with the prevailing current; for \textbf{markets}, it critiques tools that encourage contrarian gambling; for \textbf{society}, it underscores the stabilizing power of economic safety nets.
\end{abstract}

\section{Introduction}
In ancient philosophy, water is revered for its ability to adapt and overcome any obstacle not through brute force, but by yielding and flowing with the current. This wisdom, often encapsulated in the maxim "Be Water," serves as a profound metaphor for navigating the turbulent, non-stationary environment of modern financial markets. The contemporary landscape, particularly in cryptocurrencies, presents a paradox of choice. A flood of complex instruments and high-leverage tools creates an illusion of control, tempting millions of retail investors to fight the market's tide through rigid, often contrarian, strategies. This raises a fundamental question for computational social science: in a complex financial ecosystem, does survival belong to the rigid rock that stands against the current, or to the water that flows with it?

Disentangling the causal factors of retail failure in the real world is notoriously difficult. Two dominant academic narratives exist in tension. The first, rooted in behavioral finance, attributes failure primarily to endogenous cognitive biases—overconfidence, the disposition effect, and the allure of lottery-like payoffs \cite{kahneman1979prospect}. The second, from a market microstructure perspective, points to exogenous structural disadvantages—transaction costs and the insurmountable gap in initial capital \cite{piketty2014capital}. To date, no empirical study can perfectly isolate the behavioral component from the structural one.

To resolve this impasse, we construct \textbf{MAS-Utopia}, a computational laboratory designed to serve as a perfect counterfactual. Our simulation of a 10,000-agent society is intentionally engineered to eliminate all structural disadvantages. We implement a zero-friction environment and a robust Unconditional Basic Income (UBI) safety net, which reincarnates any agent whose drawdown exceeds a strict threshold with a full, equal starting capital. By creating this mathematically pristine, egalitarian environment, we can isolate the pure evolutionary fitness of trading heuristics themselves.

Our five-year, high-frequency simulation provides an unambiguous evolutionary verdict. The artificial society, through tens of thousands of cycles of bankruptcy and memetic inheritance, systematically purged strategies that attempt to fight the market's macro-tide. The overwhelmingly dominant survivor is the Trend-Following archetype, the computational embodiment of the "Be Water" principle. It does not predict the future; it simply aligns with the present, observable trend, demonstrating that in fat-tailed environments, \textit{adaptation is superior to prediction}. Based on this core finding, we bridge our simulation to reality (Sim2Real) by proposing an LLM-driven "Cognitive Prosthesis" that operationalizes this successful adaptive logic to aid real human investors.

By seamlessly integrating a multi-agent evolutionary sandbox with a real-world LLM deployment, this paper provides a holistic examination of strategic fitness. Our core contributions and implications span three critical dimensions:
\begin{itemize}
    \item \textbf{For Individual Investors:} We provide computational evidence that intuitive but flawed "buy-the-dip" strategies are evolutionarily unstable. We advocate for disciplined, low-leverage, trend-aligned strategies or, for passive participants, allocation into broad-market indices which inherently capture long-term economic trends.
    \item \textbf{For Financial Markets:} We challenge the narrative that gamified, high-frequency contrarian tools empower retail investors. Our findings advise platforms and regulators to promote products and educational resources that align with long-term, systematic trend-following, fostering "Patient Capital."
    \item \textbf{For Society and Economic Policy:} The efficacy of the UBI safety net in our simulation offers a profound insight. While it did not eliminate inequality, it crucially prevented systemic collapse and fostered a resilient, positive-sum ecosystem capable of collective learning. This underscores that robust social safety nets are not merely a mechanism for redistribution, but a precondition for a dynamic, adaptive society.
\end{itemize}

\section{Related Work}

Our research is situated at the intersection of three key domains: Agent-Based Computational Economics (ACE), the quantitative finance debate between behavioral and momentum factors, and the emerging field of LLM-driven autonomous agents.

\subsection{Agent-Based Models and Econophysics}
ACE has profoundly shifted financial modeling from representative, rational agents to heterogeneous, boundedly rational entities \cite{lebaron2001tale}. Pioneering models, such as the Santa Fe Artificial Stock Market \cite{arthur1997asset}, demonstrated that localized agent interactions spontaneously generate complex macroscopic phenomena observed in real markets, including volatility clustering and heavy-tailed return distributions. Concurrently, the field of Econophysics has used statistical mechanics to model wealth distribution, mathematically proving that even symmetric wealth exchanges can lead to Pareto-tailed distributions under certain conditions \cite{chakrabarti2013econophysics, bouchaud2008markets}. Our work extends this tradition by scaling the simulation to 10,000 agents and, more importantly, by introducing a genetic algorithm to explicitly model the \textit{evolutionary fitness} of competing trading strategies, bridging the gap between emergent phenomena and Darwinian selection.

\subsection{Behavioral Biases versus Quantitative Factors}
The debate surrounding retail investor performance is central to our research. The behavioral finance camp, led by Kahneman, Tversky \cite{kahneman1979prospect}, and empirically validated by Barber and Odean \cite{barber2000trading}, argues that cognitive biases like loss aversion and the disposition effect make retail underperformance almost inevitable. This view implicitly supports Mean-Reversion ("buy-the-dip") as an intuitive but flawed heuristic.

In stark contrast, the quantitative finance school, originating from the Fama-French three-factor model \cite{fama1993common}, has identified systematic factors that generate persistent returns. Among the most robust of these is the momentum factor, first documented by Jegadeesh and Titman \cite{jegadeesh1993returns} and later integrated into the Carhart four-factor model \cite{carhart1997persistence}. More recent work on time-series momentum \cite{moskowitz2012time} has shown that trend-following strategies are profitable across nearly all asset classes. This literature suggests that markets are not perfectly efficient and contain persistent trends. Our simulation acts as a computational arena where these two opposing worldviews—the behavioral bias towards mean-reversion and the quantitative evidence for trend-following—are pitted against each other in an evolutionary battle for survival.

\subsection{LLMs as Financial Agents and Sim2Real}
The integration of LLMs in finance has rapidly evolved from simple sentiment extraction (e.g., FinGPT \cite{zhang2023instructgpt}) to comprehensive, domain-specific models like BloombergGPT \cite{wu2023bloomberggpt}. More recently, the focus has shifted towards using LLMs as autonomous, reasoning agents capable of complex planning and tool use, as seen in the "Generative Agents" simulation \cite{park2023generative} and advanced code generation frameworks \cite{chen2024alphacodium}. Our Sim2Real application contributes to this emerging field. We do not use the LLM for prediction, but as a \textit{Cognitive Prosthesis} that operationalizes the successful Trend-Following heuristics discovered in our simulation. It translates the abstract survival laws of an artificial society into a disciplined, actionable protocol for real-world human investors, a practical application of bridging simulation with reality.

\section{Methodology: System Architecture and Micro-Mechanisms}

To rigorously evaluate the micro-mechanisms of wealth depletion and the emergence of stratification, we formalize the \textbf{MAS-Utopia} environment. The simulation processes 5 years of 5-minute interval OHLCV (Open, High, Low, Close, Volume) data across an asset universe $\mathcal{C}$ where $|\mathcal{C}| = 100$. This yields a continuous temporal grid of over $5.2 \times 10^5$ steps. To ensure the simulation captures the true mathematical boundaries of financial markets while remaining computationally tractable, we engineer several institutional-grade micro-mechanisms.

\subsection{Agent Genomic Architecture and Behavioral Archetypes}
The artificial society is populated by $N = 10,000$ heterogeneous agents. To avoid the ''black box'' nature of purely randomized neural networks, we utilize an interpretable, heuristic-based genomic architecture. Each agent $i$ operates under a specific trading archetype $\mathcal{S}_i \in \{Sniper, Trend, Reversion, Grid, HFT\}$, which dictates their prior parameter distributions.

The operational boundaries of an agent are defined by a high-dimensional continuous genetic vector $\mathbf{G}_i$:
\begin{equation}
    \mathbf{G}_i = \langle l_i, s_i, \theta_{SL}^{(i)}, \theta_{TP}^{(i)}, \phi_i, \rho_i, \mathcal{W}_i \rangle
\end{equation}

Each parameter corresponds to a distinct behavioral finance concept:
\begin{itemize}
    \item \textbf{Leverage and Position Sizing ($l_i, s_i$):} $l_i \in[1, 100]$ denotes the leverage multiplier. $s_i \in (0, 1]$ represents the fraction of available equity deployed per trade, acting as a proxy for the Kelly Criterion risk appetite. For instance, \textit{Sniper} agents exhibit $l_i \to \max$ but $s_i \to \min$, simulating asymmetric "lottery-ticket" betting behaviors.
    \item \textbf{Risk Tolerance Multipliers ($\theta_{SL}^{(i)}, \theta_{TP}^{(i)}$):} These define the Stop-Loss and Take-Profit widths. \textit{Grid} agents possess $\theta_{SL} \gg \theta_{TP}$, embodying disposition-effect behaviors (holding losers, selling winners early).
    \item \textbf{Cognitive Bias Factors ($\phi_i, \rho_i$):} Parameterized in $[0, 1]$, $\phi_i$ quantifies FOMO (Fear Of Missing Out) and $\rho_i$ quantifies Panic. These factors map to Kahneman's Prospect Theory, determining how an agent's logic is overridden during extreme market anomalies.
    \item \textbf{Localized Attention ($\mathcal{W}_i$):} $\mathcal{W}_i \subseteq \mathcal{C}$ defines the agent's specific capability circle (e.g., focusing only on large-cap vs. highly volatile meme assets), simulating the bounded attention span of human retail investors.
\end{itemize}

\subsection{Adaptive Volatility Anchoring: Non-Stationary Risk Control}
A critical flaw in naive retail simulations is the reliance on fixed-percentage stop-loss thresholds (e.g., a static 5\% cut-off). Financial time series exhibit volatility clustering and fat-tailed distributions (Mandelbrot effects), rendering static thresholds highly susceptible to being prematurely triggered by localized noise.

To address this, we introduce a dynamic \textit{Average True Range (ATR)} anchoring mechanism. The True Range at time $t$ for asset $c$ is calculated as the maximum of three variance measures:
\begin{equation}
    TR_{c,t} = \max(H_t - L_t, |H_t - C_{t-1}|, |L_t - C_{t-1}|)
\end{equation}
For any opened position with an entry price $P_{entry}$, the execution thresholds are dynamically scaled by the smoothed ATR ($\sigma_{c,t}^{ATR}$) and the agent's genetic tolerance:
\begin{equation}
    SL_{i, c, t} = 
    \begin{cases} 
      P_{entry} - (\sigma_{c,t}^{ATR} \times \theta_{SL}^{(i)}) & \text{if Long} \\
      P_{entry} + (\sigma_{c,t}^{ATR} \times \theta_{SL}^{(i)}) & \text{if Short}
    \end{cases}
\end{equation}
This micro-mechanism ensures that agents adapt to changing market regimes: during violent macroeconomic shocks, their operational boundaries automatically widen to absorb noise, preventing cascading liquidation feedback loops within the simulation.

\subsection{Hierarchical Regime Filtering and Heuristic Decision Engine}
Boundedly rational agents do not trade at random; their decisions emerge from a confluence of macroscopic trends and microscopic signals. We model this via a multi-factor heuristic scoring function $\mathbb{S}_{i, c, t} \in [-100, 100]$.

First, a \textit{Macroscopic Regime Filter} ($\mathcal{M}_t$) is established by evaluating the momentum of the market beta (e.g., Bitcoin dominance):
\begin{equation}
    \mathcal{M}_t = 
    \begin{cases} 
      BULL & \text{if } RSI_{BTC, t} > 60 \\
      BEAR & \text{if } RSI_{BTC, t} < 40 \\
      FLAT & \text{otherwise}
    \end{cases}
\end{equation}

Subsequently, the localized score aggregates technical states and psychological biases. For instance, consider an asset exhibiting severe overbought conditions ($RSI_{c,t} > 85$). The scoring logic diverges based on the agent's archetype $\mathcal{S}_i$ and cognitive genes:
\begin{equation}
    \mathbb{S}_{i, c, t} \mathrel{+}= 
    \begin{cases} 
      +10 \cdot \phi_i & \text{if } \mathcal{S}_i = Trend \text{ (Irrational Exuberance)} \\
      -15 \cdot (1-\phi_i) & \text{if } \mathcal{S}_i = Reversion \text{ (Contrarian Short)}
    \end{cases}
\end{equation}
Execution is triggered strictly when the absolute conviction exceeds a hard threshold: $|\mathbb{S}_{i, c, t}| \geq \Omega$. This hierarchical design effectively simulates the friction between an investor's strategic discipline and their emotional impulses.

\subsection{Market Microstructure Constraints: The Liquidity Shackle}
A persistent methodological illusion in quantitative backtesting is the "Infinite Compounding Fallacy"—the assumption that a compounding agent can execute infinite size without impacting the order book. In reality, market depth strictly limits capital utilization.

We resolve this by enforcing a \textit{Whale Liquidity Shackle} ($\mathcal{L}_{max} = \$500,000$). The actual notional value $\mathcal{V}_{i,c,t}$ deployed by an agent with equity $E_{i,t}$ is mathematically bounded:
\begin{equation}
    \mathcal{V}_{i, c, t} = \min \Big( E_{i,t} \cdot s_i \cdot l_i, \;\; \mathcal{L}_{max} \Big)
\end{equation}
Consequently, the required isolated margin is frozen as $\mathcal{M}_{used} = \mathcal{V}_{i, c, t} / l_i$. This mechanism acts as a critical structural dampener. As an agent becomes ultra-wealthy, their capital utilization rate plummets asymptotically toward zero. This accurately mirrors institutional liquidity constraints and prevents single agents from generating unrealistic trillions of dollars in the simulation, ensuring the validity of our Gini coefficient measurements.

\subsection{Endogenous Margin Calculus and the Leverage-Tolerance Constraint}
A fundamental mechanism often overlooked in simplified ACE models is the endogenous mathematical coupling between an agent's leverage ($l_i$) and their stop-loss tolerance ($\theta_{SL}^{(i)}$). In our simulation, isolated margin mechanics are strictly enforced to replicate real-world clearinghouse constraints.

For a deployed notional value $\mathcal{V}_{i, c, t}$, the required frozen margin is $\mathcal{M}_{used} = \mathcal{V}_{i, c, t} / l_i$. A position faces forced liquidation (margin call) if the floating loss entirely depletes this margin. Therefore, the absolute maximum price deviation $\Delta P_{liq}$ an agent can endure before systemic ruin is inversely proportional to their leverage:
\begin{equation}
    \Delta P_{liq} \approx \frac{P_{entry}}{l_i}
\end{equation}

This introduces a critical survival constraint. For an agent's genetic Stop-Loss (derived from $ATR$ via Equation 3) to function as a protective cognitive boundary rather than a post-mortem artifact, the physical stop-loss distance must be strictly tighter than the exchange's liquidation distance:
\begin{equation}
    \left( \sigma_{c,t}^{ATR} \times \theta_{SL}^{(i)} \right) < \frac{P_{entry}}{l_i}
\end{equation}

This mathematically dictates the evolutionary viability of different archetypes. For instance, the \textit{Sniper} archetype operates at extreme leverage ($l_i \to 100$). Equation 8 dictates that their liquidation distance is a mere $1\%$. Consequently, Snipers are evolutionarily forced to maintain an extremely narrow stop-loss tolerance ($\theta_{SL} \to 0.2$). Conversely, \textit{Grid} and \textit{Reversion} agents utilize low leverage ($l_i \le 5$), granting them a wide liquidation boundary ($20\%$), which mathematically permits high $\theta_{SL}$ values, allowing them to absorb short-term noise and ride out mean-reverting volatility.

\subsection{Market Microstructure Constraints: The Liquidity Shackle}
Another persistent methodological illusion in quantitative backtesting is the "Infinite Compounding Fallacy"—the assumption that a compounding agent can execute infinite size without impacting the order book. In reality, market depth strictly limits capital utilization.

We resolve this by enforcing a \textit{Whale Liquidity Shackle} ($\mathcal{L}_{max} = \$500,000$). The actual notional value $\mathcal{V}_{i,c,t}$ deployed by an agent with equity $E_{i,t}$ and Kelly-sizing $s_i$ is mathematically bounded:
\begin{equation}
    \mathcal{V}_{i, c, t} = \min \Big( E_{i,t} \cdot s_i \cdot l_i, \;\; \mathcal{L}_{max} \Big)
\end{equation}
This mechanism acts as a critical structural dampener. As an agent becomes ultra-wealthy, their capital utilization rate ($U_i = \mathcal{V}_{i} / E_{i}$) plummets asymptotically toward zero. This accurately mirrors institutional liquidity constraints and prevents single agents from generating unrealistic trillions of dollars in the simulation, ensuring the absolute validity of our Gini coefficient measurements.

\subsection{Egalitarian UBI Restart and Memetic Evolution}
To strictly isolate endogenous behavioral failures from exogenous institutional inequality, we enforce an absolute egalitarian Unconditional Basic Income (UBI) axiom. Social death (bankruptcy) is triggered if an agent's Maximum Drawdown exceeds a strict threshold: $DD_{i,t} \geq \Gamma_{ruin}$ (where $\Gamma_{ruin} = 0.40$).

Upon social death, the agent is purged and regenerated via a genetic algorithm. Crucially, to maintain the "Utopian" baseline, \textbf{no wealth is inherited}. Every newborn agent—regardless of its genetic origin—is instantiated with an identical, perfectly equal starting capital $C_0 = \$10,000$. The evolution occurs purely at the \textit{memetic} (informational) level via a bifurcated probability path ($\alpha = 0.5$):
\begin{itemize}
    \item \textbf{Epistemic Inheritance (Probability $\alpha$):} The newborn inherits the strategy archetype $\mathcal{S}$ and genetic vector $\mathbf{G}$ of a Top-10 elite survivor. To simulate imperfect imitation of successful strategies, the vector is subjected to Gaussian mutation: $\mathbf{G}_{new} = \mathbf{G}_{elite} + \mathcal{N}(0, \sigma^2)$. 
    \item \textbf{Randomized Generation (Probability $1-\alpha$):} The agent is initialized with a completely randomized archetype and genetic matrix, ensuring continuous exploration of the strategy parameter space.
\end{itemize}

By guaranteeing that $E_{i, t_{born}} \equiv \$10,000$ for all agents across all generations, we construct a mathematically pristine sandbox. Any emergence of severe wealth stratification (high Gini coefficient) in this environment can be undeniably attributed to the compounding mathematical nature of leveraged trading and behavioral variance, rather than historical capital accumulation.

\section{Experimental Results and Discussion}

Our 5-year simulation of a 10,000-agent society generated a rich dataset reflecting the economic and sociological evolution of an artificial civilization. The findings robustly support a narrative of collective learning and adaptation, culminating in a positive-sum outcome driven by the evolutionary triumph of a single strategic archetype.

\subsection{Emergence of a Positive-Sum Economy}
\begin{figure}[h!]
    \centering
    \includegraphics[width=0.9\textwidth]{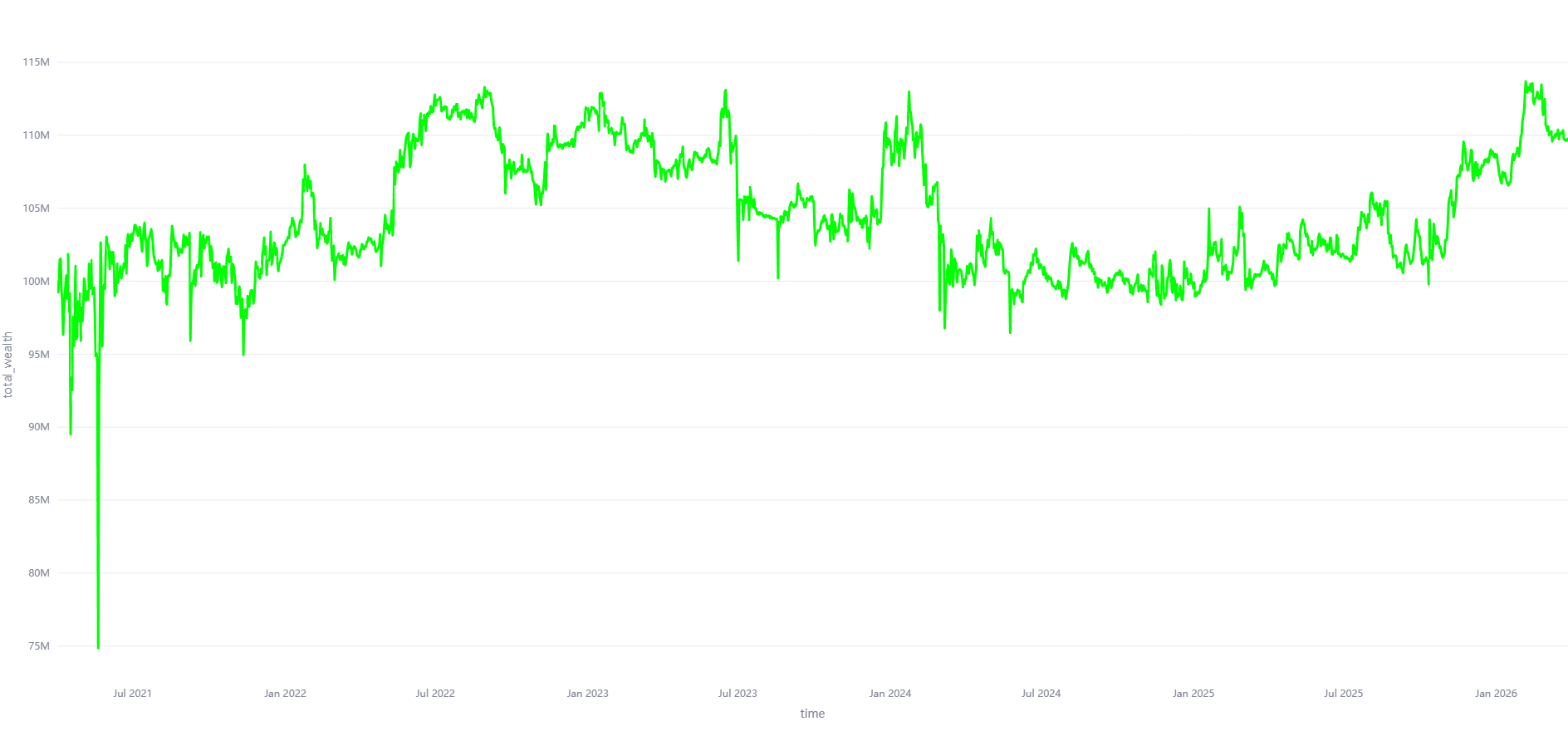}
    \caption{Aggregate wealth of the 10,000-agent society over 5 years. The net positive growth indicates a successful positive-sum outcome.}
    \label{fig:total_wealth}
\end{figure}

A primary and counter-intuitive finding is that the society, as a collective, achieved a positive-sum outcome. As illustrated in Figure \ref{fig:total_wealth}, the aggregate wealth grew from \$100M to over \$110M. 
\textbf{Deep Dive:} This phenomenon refutes the assumption that retail-dominated markets are inherently zero-sum. The initial dip in 2021 can be interpreted as a societal "tuition fee," where a high volume of agents with maladaptive strategies (e.g., high leverage, contrarian betting in a bull market) were rapidly purged. The subsequent recovery and growth indicate that the evolutionary mechanism successfully propagated survival-positive traits, enabling the collective to systematically extract value from the market's inherent, long-term upward drift (beta) and from the capital of liquidated, misaligned agents.

\subsection{Ecological Succession and the Triumph of Trend-Following}
\begin{figure}[h!]
    \centering
    \includegraphics[width=\textwidth]{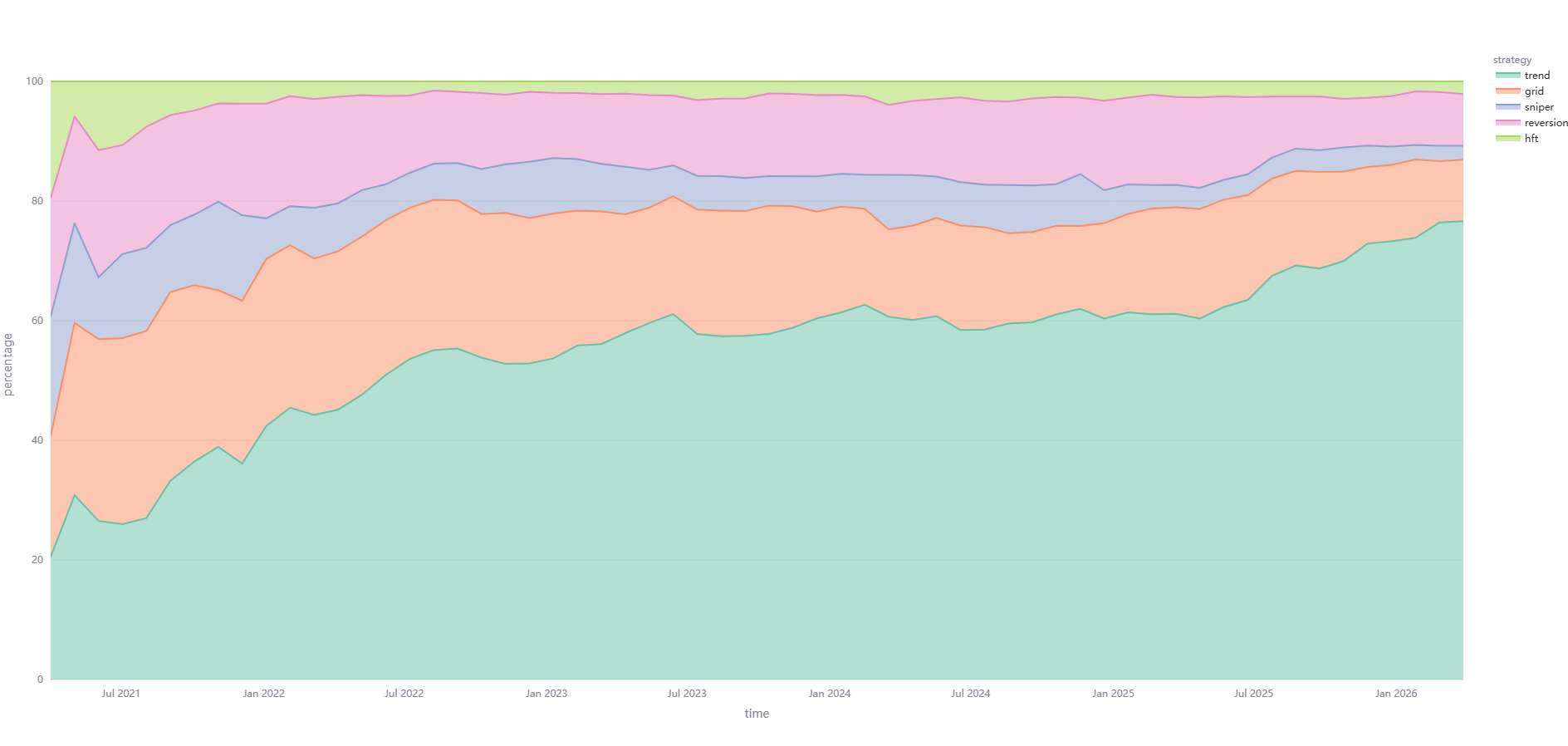}
    \caption{Ecological succession of the five trading archetypes. The Trend-Following strategy (light green) exhibits clear evolutionary dominance.}
    \label{fig:river_plot}
\end{figure}

The Ecological Succession Chart (Figure \ref{fig:river_plot}) provides the central narrative of our experiment. While all five archetypes began with equal population shares, a dramatic evolutionary convergence occurred.
\textbf{Deep Dive:} This succession is not random; it is a textbook case of natural selection driven by return-distribution asymmetry. Mean-Reversion and Grid strategies exhibit a \textit{negative skew}: they generate many small, consistent profits during range-bound markets but suffer rare, catastrophic losses during strong, "fat-tailed" trends. This is analogous to selling out-of-the-money options. In contrast, the Trend-Following strategy exhibits a \textit{positive skew}: it endures many small, manageable losses during consolidations but captures explosive, outsized gains during macro-directional moves. It is structurally designed to profit from "black swan" events. The market, through its periodic trending phases, acted as an evolutionary filter, systematically selecting for positive skew and against negative skew, leading to the near-total dominance of the Trend-Following phenotype (Figure \ref{fig:pie_chart}).

\subsection{Quantitative Analysis of Archetype Fitness}
A quantitative breakdown of the surviving population (Table \ref{tab:strategy_stats}) reinforces this narrative with stark clarity.
\textbf{Deep Dive:} The Trend-Following cohort is the only archetype to achieve a significant positive average return (+14.71\%), despite having the \textit{lowest} average trade count (312 trades). This reveals a crucial insight: in fat-tailed markets, profitability is inversely correlated with trading frequency. Hyperactive strategies (e.g., Sniper, HFT) incur higher "volatility tax" (whipsaw losses) and are more likely to be active during unfavorable regimes. The low-frequency nature of the Trend-Following strategy acts as a natural filter, forcing it to participate only in high-conviction, macro-scale moves, thus maximizing its positive-skew advantage.

\begin{table}[h!]
\centering
\caption{Performance Statistics of Surviving Archetypes at Simulation End}
\label{tab:strategy_stats}
\begin{tabular}{@{}lrrrrr@{}}
\toprule
\textbf{Strategy Type} & \textbf{Survivors} & \textbf{Avg. Asset (\$)} & \textbf{Avg. ROI (\%)} & \textbf{Avg. Trades} \\ \midrule
trend                  & 7,659              & 11,470.95                & +14.71                 & 312                  \\
reversion              & 865                & 10,038.46                & +0.38                  & 1,483                \\
hft                    & 213                & 9,937.89                 & -0.62                  & 1,277                \\
grid                   & 1,030              & 9,530.52                 & -4.69                  & 1,254                \\
sniper                 & 233                & 9,608.63                 & -3.91                  & 1,665                \\ \bottomrule
\end{tabular}
\end{table}

\subsection{Capped Inequality and The Rise of a Resilient Elite}
\begin{figure}[h!]
    \centering
    \includegraphics[width=0.9\textwidth]{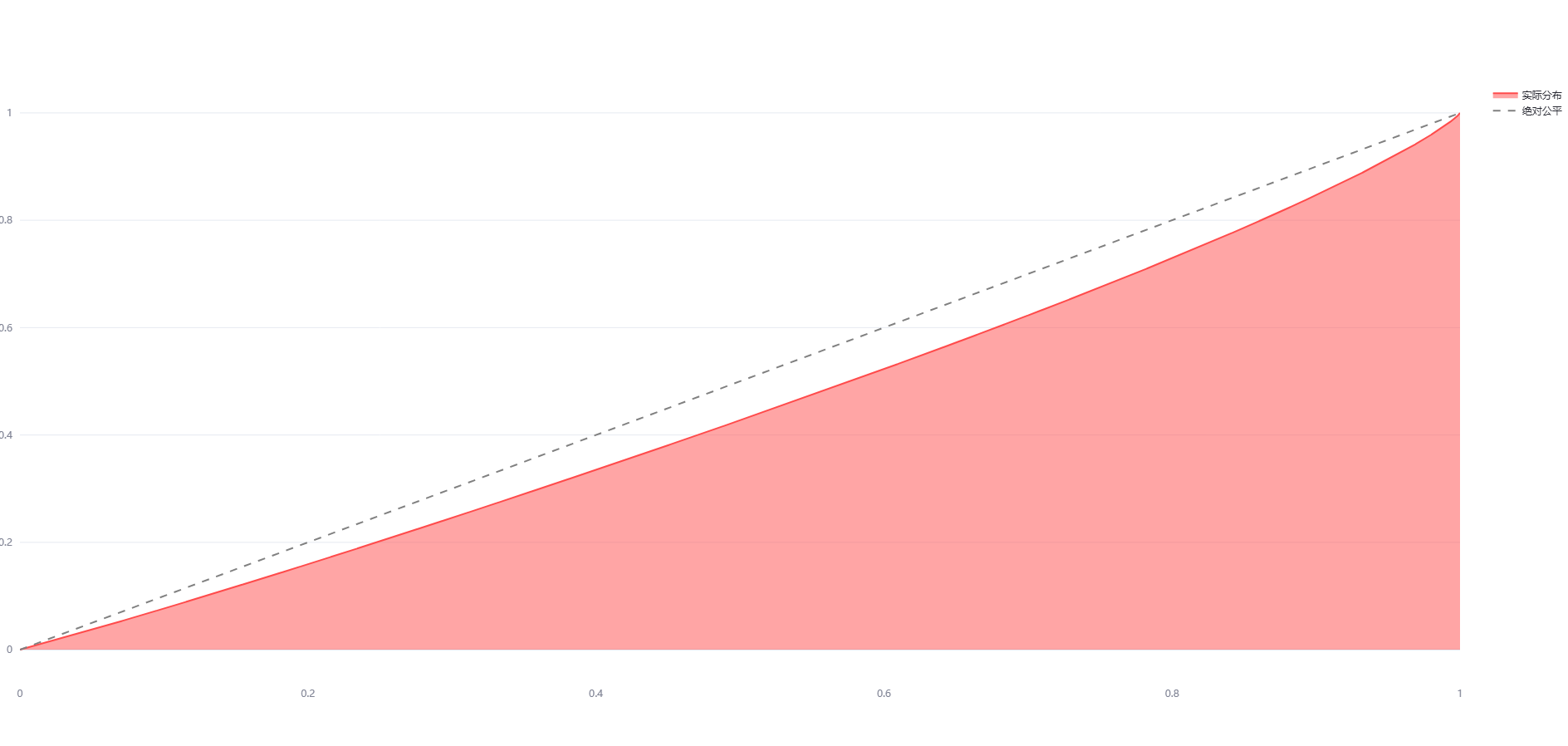} 
    \caption{Lorenz Curve of all agents.}
    \label{fig:fig_lorenz_curve}
\end{figure}

\begin{figure}[h!]
    \centering
    \includegraphics[width=0.9\textwidth]{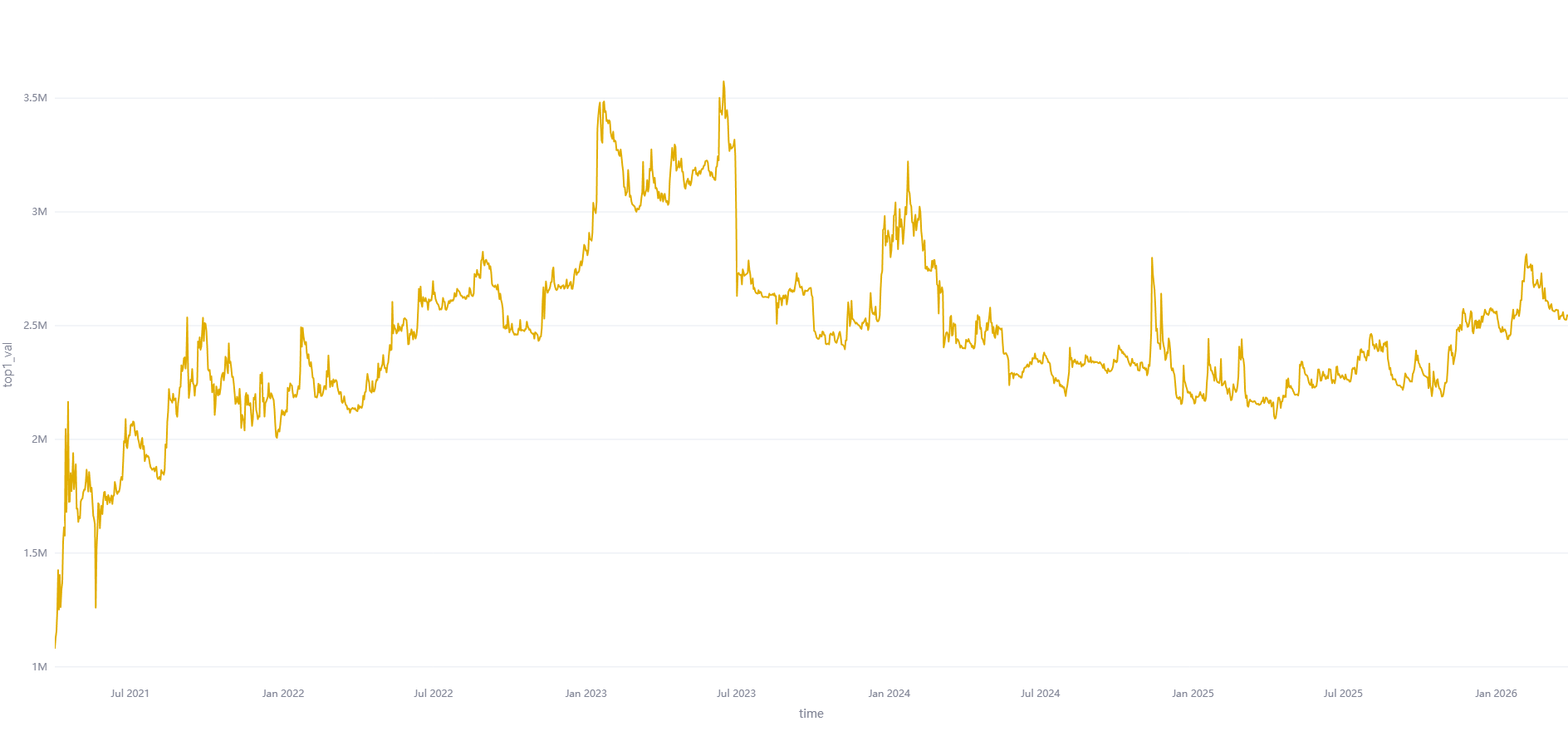} 
    \caption{Total wealth held by the Top 1\% of the population over 5 years. This index serves as a proxy for the health and adaptability of the elite class.}
    \label{fig:top1_wealth}
\end{figure}

\begin{figure}[h!]
    \centering
    \includegraphics[width=0.48\textwidth]{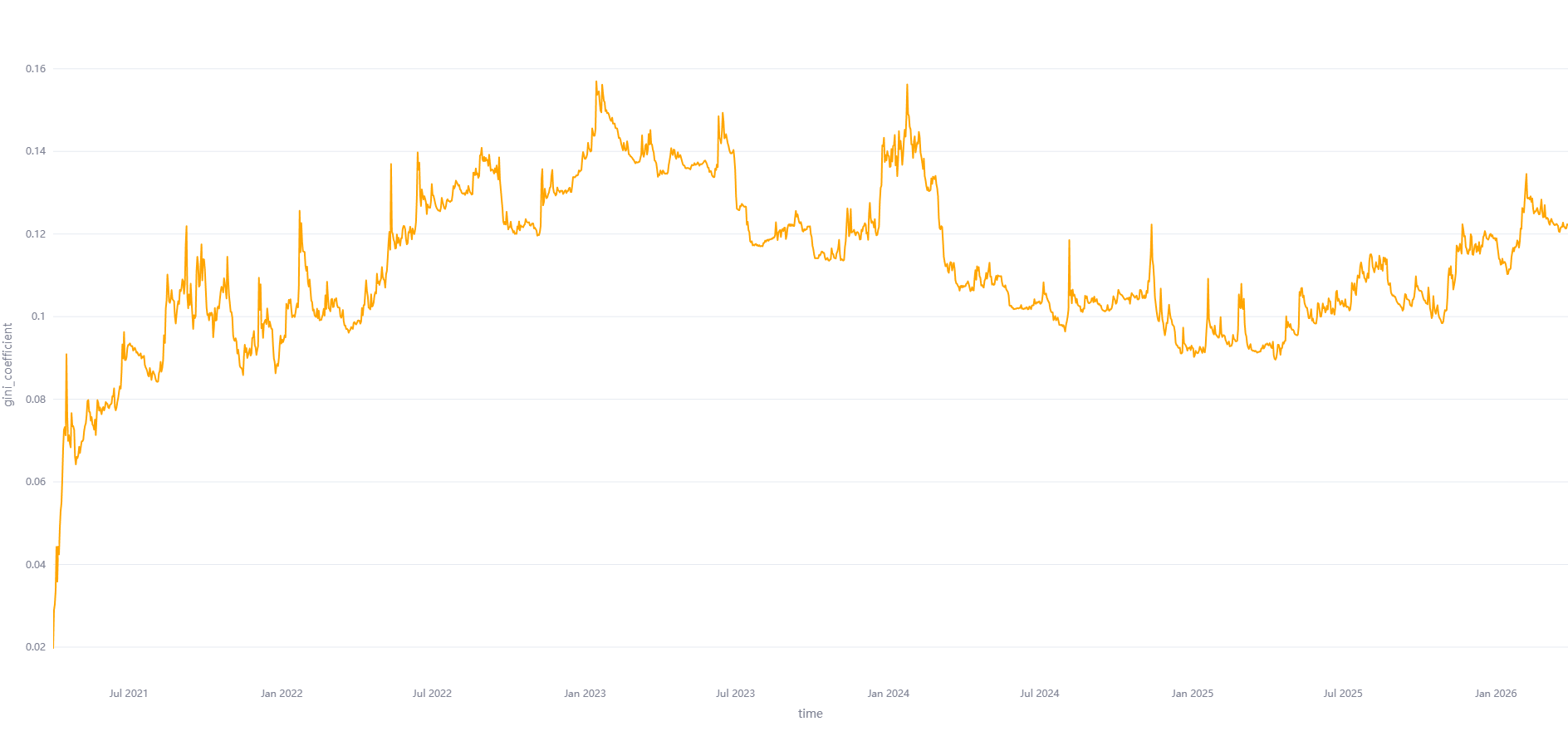}
    \hfill
    \includegraphics[width=0.48\textwidth]{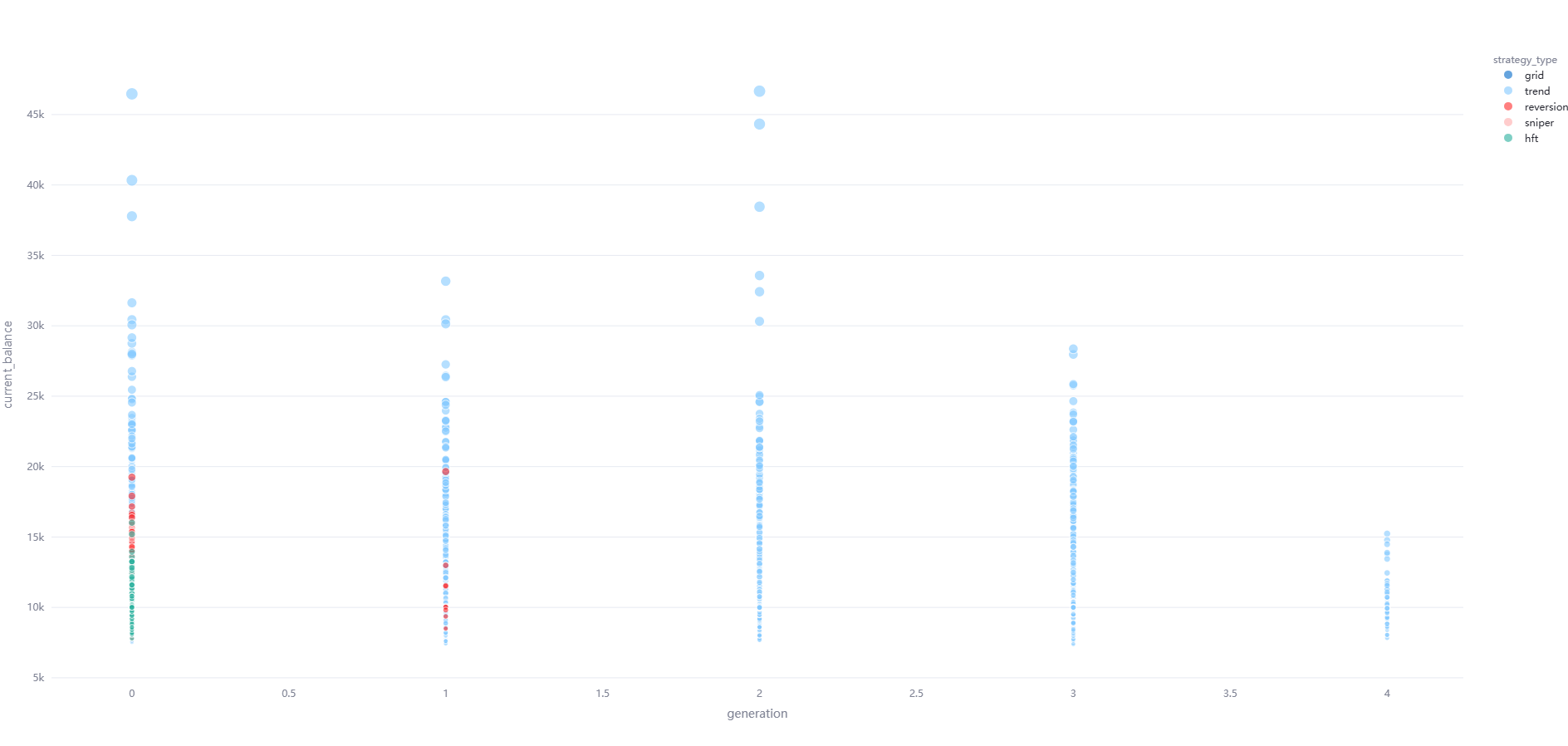}
    \caption{Left: Gini coefficient stabilizing. Right: Upward wealth mobility across generations.}
    \label{fig:gini_curve}
    \label{fig:generational_wealth}
\end{figure}

The Gini Coefficient (Figure \ref{fig:gini_curve}) and Lorenz Curve (Figure \ref{fig:fig_lorenz_curve}) provide a macroscopic view of social structure, demonstrating that the UBI safety net successfully prevented a "winner-take-all" monopolistic state. However, the trajectory of the Top 1\% wealth index (Figure \ref{fig:top1_wealth}) reveals a more nuanced and profound narrative about the formation of a resilient, adaptive elite class.

\textbf{Deep Dive - Phase 1 (2021-mid 2022): The Great Filter.} The initial phase is characterized by extreme volatility in the elite's wealth. This reflects a period of intense ecological competition. The 2021 bull market allowed a diverse range of archetypes—including high-leverage Snipers—to temporarily reach the top 1\%, creating a fragile and unstable elite. The subsequent market crash acted as a "Great Filter," systematically purging these maladaptive, high-risk phenotypes from the upper echelons of society.

\textbf{Deep Dive - Phase 2 (mid 2022 - mid 2023): The Momentum Consolidation.} Following the Great Filter, the elite class undergoes a consolidation. The wealth index begins a steady, less volatile ascent, peaking in mid-2023. This phase corresponds directly to the "Trend-Following Singularity" observed in the ecological succession plot. The elite class is no longer a random collection of lucky gamblers; it has become a homogenous cohort dominated by agents who have mastered the art of surviving bear markets and capturing macro trends. Their collective wealth grows as they absorb the capital of the continuously failing lower classes.

\textbf{Deep Dive - Phase 3 (mid 2023 onwards): Dynamic Equilibrium and Generational Learning.} The final phase shows the elite's wealth oscillating within a stable, high-value range. It no longer grows exponentially. This indicates the system has reached a dynamic equilibrium. The elite Trend-Followers are efficiently extracting value, but the UBI mechanism ensures a constant influx of new, well-capitalized "challengers," preventing the elites from achieving absolute monopoly. Furthermore, the generational wealth chart (Figure \ref{fig:generational_wealth}) proves that this elite status is not static; it is passed down through successful memetic inheritance. Later generations of Trend-Followers are born with more refined genetic parameters, allowing them to consistently challenge and replace older elites, ensuring the continued fitness and prosperity of the society as a whole. This is not a story of a fixed aristocracy, but of a meritocratic, constantly evolving class of survivors.

\subsection{The Genetic Footprint of Survival}
\begin{figure}[h!]
    \centering
    \includegraphics[width=0.48\textwidth]{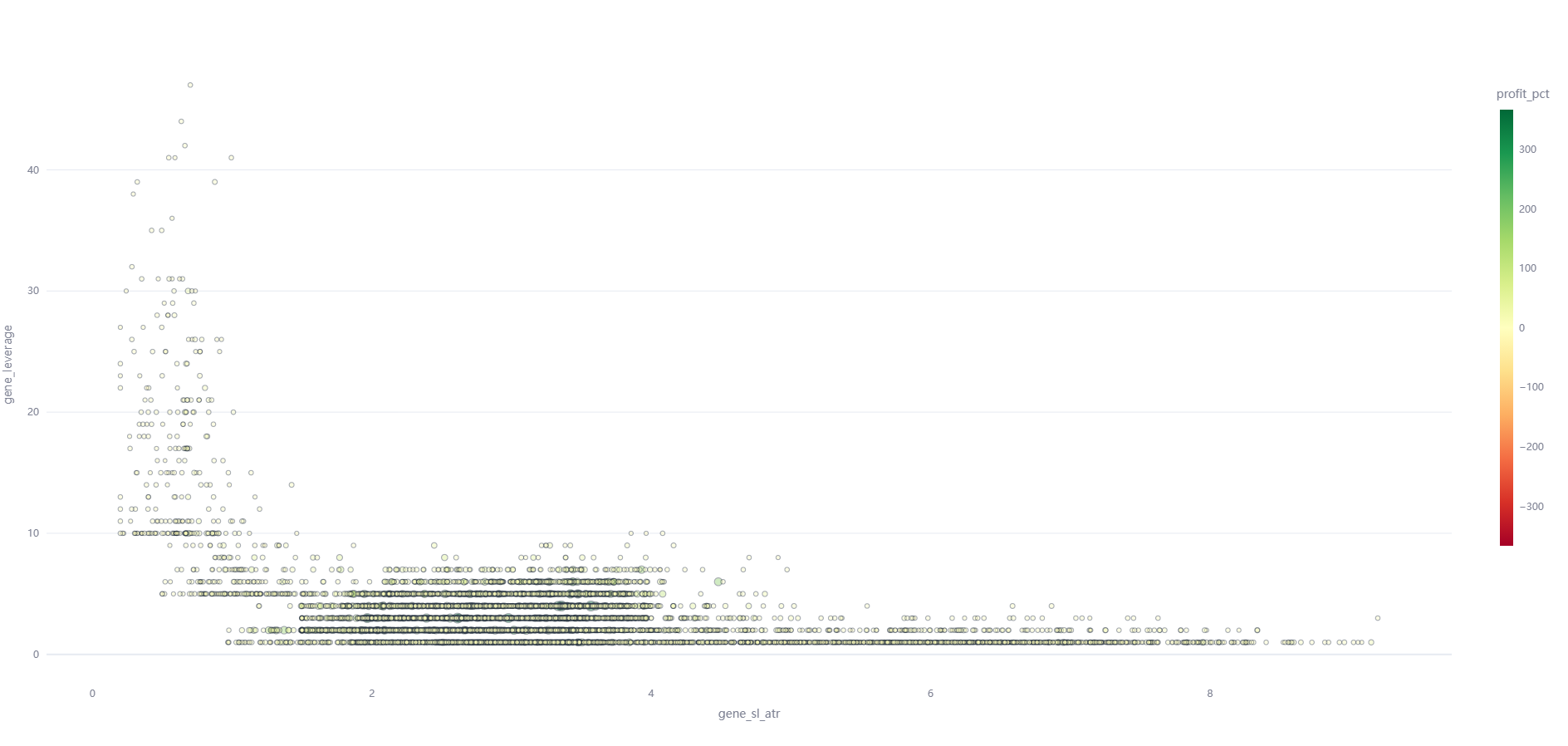}
    \hfill
    \includegraphics[width=0.48\textwidth]{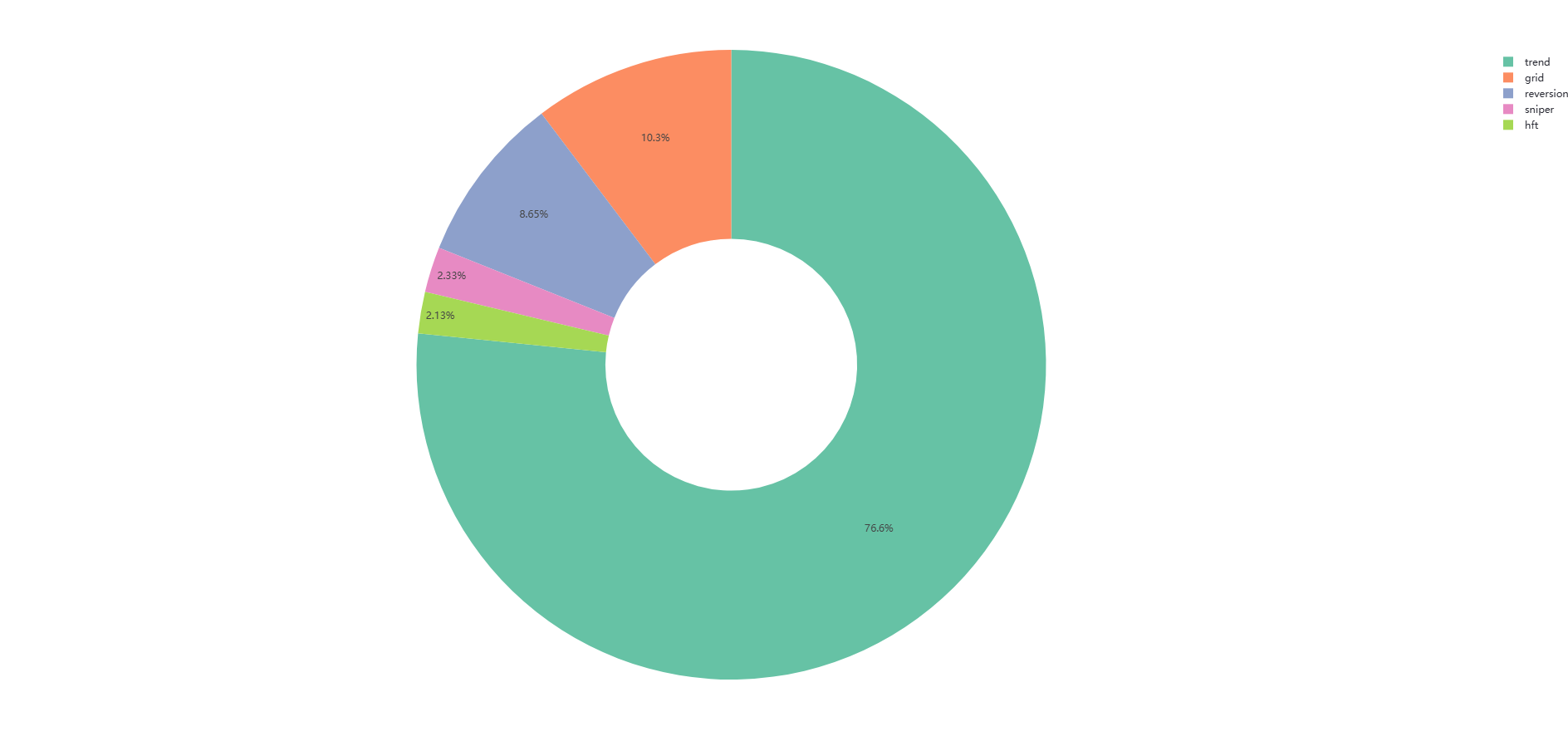}
    \caption{Left: Genotype heatmap of surviving agents. Right: Final population distribution.}
    \label{fig:gene_heatmap}
    \label{fig:pie_chart}
\end{figure}

\begin{figure}[h!]
    \centering
    \includegraphics[width=0.9\textwidth]{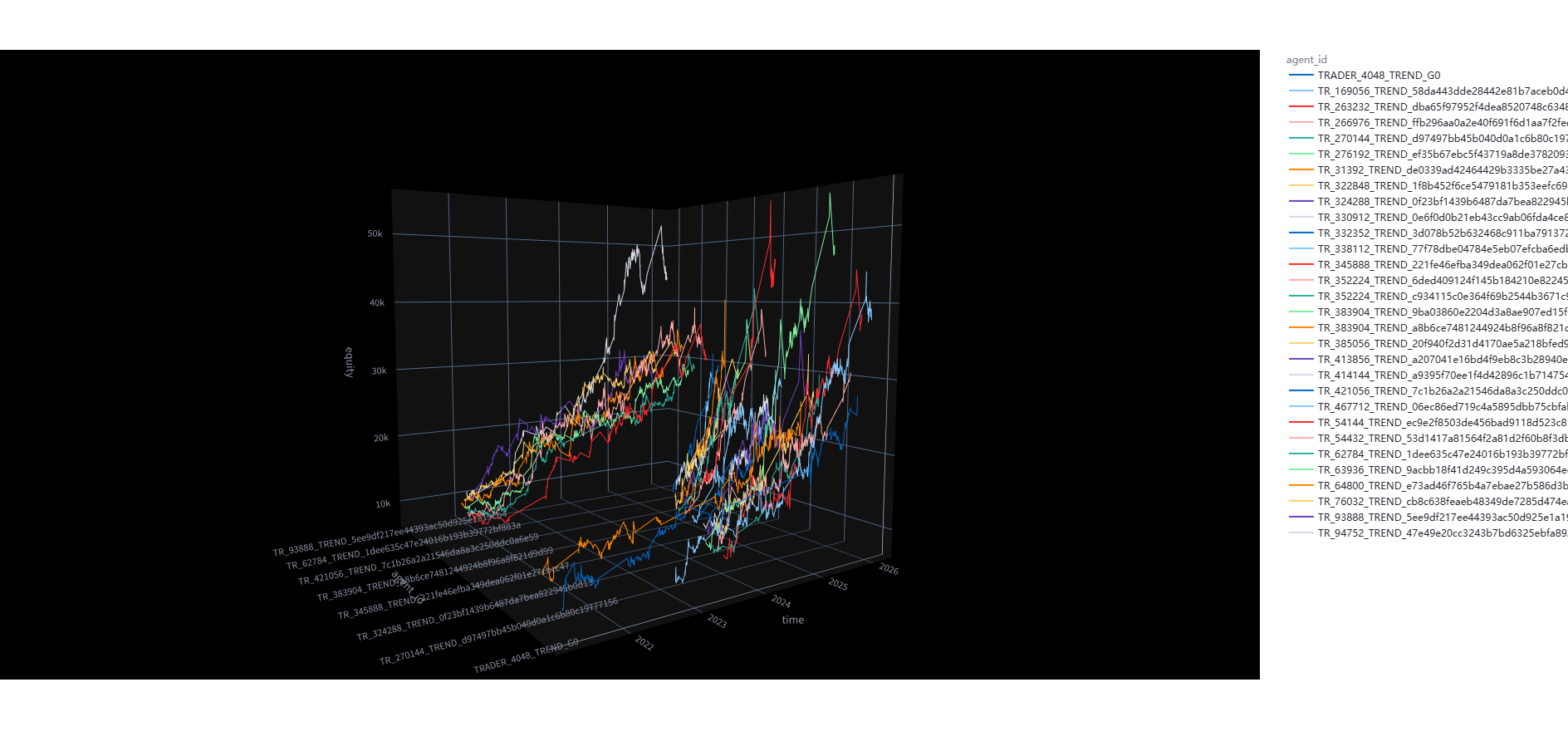}
    \caption{3D wealth trajectories of the Top 30 elite agents over time.}
    \label{fig:3d_trajectory}
\end{figure}

The agent genotype heatmap (Figure \ref{fig:gene_heatmap}) serves as a micro-level validation of our macro findings.
\textbf{Deep Dive:} The profitable agents (green dots) are overwhelmingly concentrated in the low-leverage ($l_i < 10$) and moderate stop-loss ($1 < \theta_{SL} < 4$) region. This forms a "Survival Zone" in the genetic parameter space. Agents outside this zone, particularly those in the high-leverage quadrant, exhibit almost universally negative returns. This provides a clear, data-driven prescription for risk management: success is not about maximizing leverage, but about optimizing survival. The 3D trajectory plot (Figure \ref{fig:3d_trajectory}) further visualizes the compounding paths of the top elite, who are almost exclusively high-generation Trend-Followers operating within this Survival Zone.

\clearpage

\section{Sim2Real: An LLM-Driven Trend Confirmation System}

The core lesson from MAS-Utopia is that survival in fat-tailed markets requires a disciplined, hierarchical decision process: first, identify a macro-level anomaly, and second, await a micro-level confirmation of a trend shift before committing capital. To bridge this finding with reality, we engineered and deployed a production-grade, LLM-driven "Cognitive Prosthesis." This system operationalizes the successful agent's logic to serve real human investors, acting as an emotional dampener and a rational anchor. The system's architecture comprises three distinct, interconnected layers.

\subsection{Layer 1: Global Anomaly Radar (Macro-Scanner)}
Analogous to an agent scanning the market for opportunities, this layer performs a high-concurrency scan across the entire US and HK equity universes. It acts as a heuristic dimensionality reduction engine, filtering thousands of assets down to a manageable list of candidates exhibiting potential macro-dislocation (e.g., price significantly below the 200-day moving average and RSI indicating oversold conditions). To ensure robustness against real-world network failures, this layer implements a multi-tier fallback topology, dynamically downgrading data acquisition across multiple financial APIs if rate limits are encountered.

\subsection{Layer 2: Watchlist MACD Trigger Sentinel (Micro-Confirmation)}
\begin{figure}[h!]
    \centering
    \includegraphics[width=0.8\textwidth]{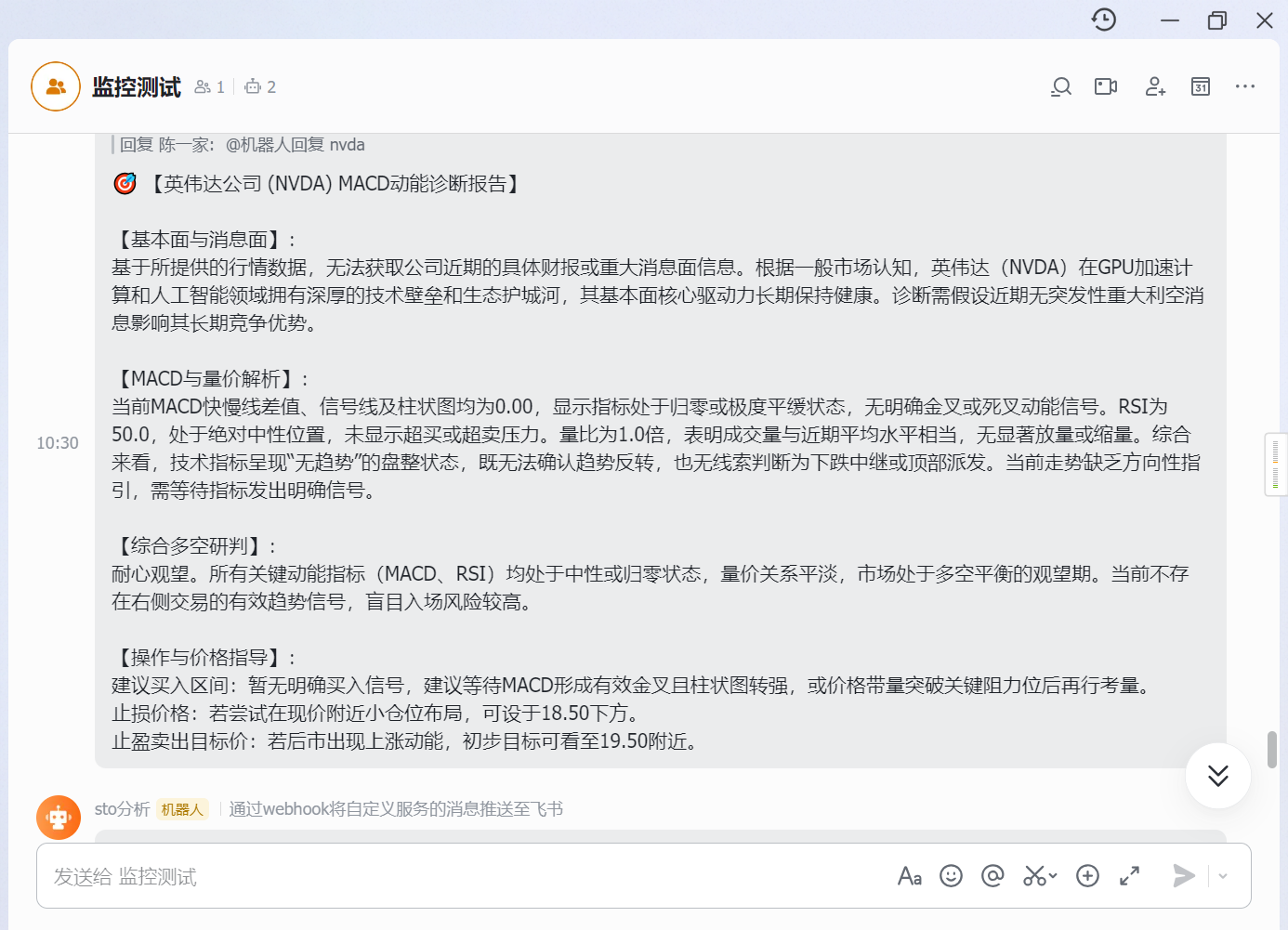}
    \caption{The Layer 2 Radar output, delivered via IM. It identifies a single asset with a MACD golden cross and provides a preliminary AI score, demonstrating the system's filtering and anti-bombardment capabilities.}
    \label{fig:radar}
\end{figure}

Once an asset enters the watchlist, the second layer takes over. This stateful sentinel's single, disciplined mission is to detect the specific trend-reversal signal that proved evolutionarily superior in our simulation: the \textbf{MACD golden cross} or other significant volume/RSI surges. This layer acts as the critical confirmation filter, preventing premature "knife-catching." Furthermore, to prevent alert fatigue for the human user, we implement an intelligent "anti-bombardment" mechanism. As shown in Figure \ref{fig:radar}, the system compares the current alert cohort with the previously pushed cohort; if they are identical, the notification is suppressed, respecting the user's attention span.

\subsection{Layer 3: LLM-Powered Tactical Report Generation (Execution Protocol)}
\begin{figure}[h!]
    \centering
    \includegraphics[width=0.8\textwidth]{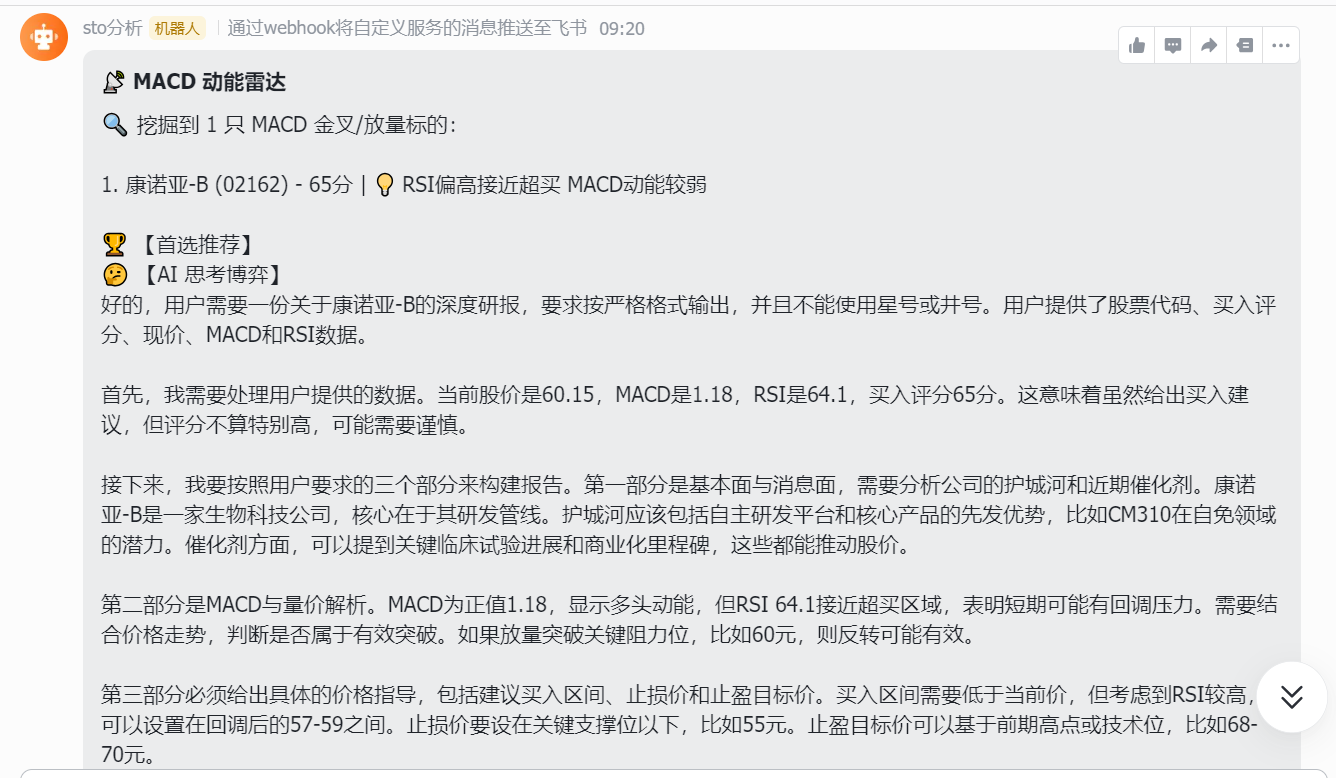}
    \caption{Explicit extraction of the LLM's Chain-of-Thought reasoning. This provides the user with transparent, interpretable insight into the AI's decision-making process before presenting the final conclusion.}
    \label{fig:deepseek_cot}
\end{figure}

When the Layer 2 sentinel fires a confirmed alert, the final layer is activated. The ticker and its data are injected into a Large Language Model (e.g., DeepSeek-Reasoner), which is constrained by a strict "Quantitative Analyst" persona prompt. A key innovation in our architecture is the explicit extraction of the LLM's Chain-of-Thought (CoT) reasoning process. As shown in Figure \ref{fig:deepseek_cot}, the system programmatically separates the LLM's internal, step-by-step logical deduction  from its final, polished report.

\begin{figure}[h!]
    \centering
    \includegraphics[width=0.8\textwidth]{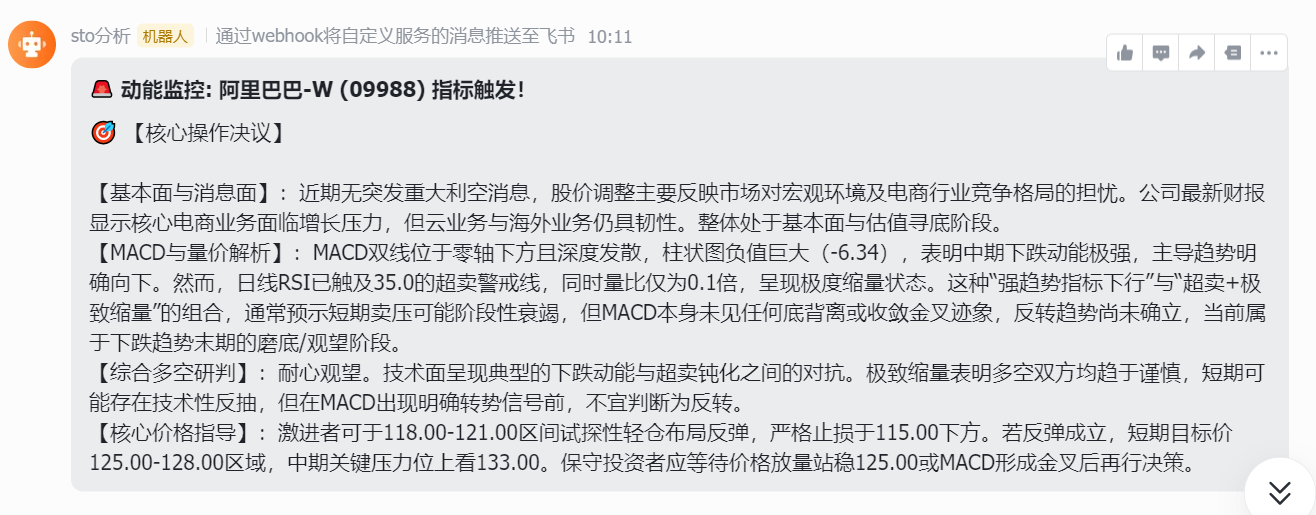}
    \caption{The final, structured tactical report delivered to the user. It provides a holistic analysis and a concrete, risk-managed trading plan, directly translating the simulation's survival laws into an actionable format.}
    \label{fig:final_report}
\end{figure}

This separated reasoning is first delivered to the user, providing full transparency. Subsequently, the final, structured tactical report is generated and pushed (Figure \ref{fig:final_report}). This report synthesizes fundamental context with a rigorous analysis of the MACD signal's validity, and most importantly, provides a concrete, risk-managed execution plan, including specific price levels for entry, stop-loss, and take-profit. This completes the Sim2Real loop: a principle discovered in a massive artificial society is delivered as a disciplined, transparent, and actionable protocol to a real human investor, helping them overcome the very biases that led to the extinction of other archetypes in our simulation.

\clearpage

\section{Conclusion and Interdisciplinary Implications}

In this paper, we presented MAS-Utopia, a large-scale, multi-agent financial simulation designed to uncover the evolutionarily stable strategies in a complex, volatile market. By establishing a zero-friction, UBI-backed counterfactual baseline, we isolated the impact of trading heuristics from structural inequalities. Our five-year simulation yielded a clear and powerful result: the artificial society achieved a collective positive-sum outcome, driven by the overwhelming evolutionary triumph of the Trend-Following archetype. This finding challenges the prevailing narrative of inevitable retail failure and provides a data-driven blueprint for survival based on disciplined, macro-aligned strategies.

Our core contributions and their broader implications span three critical dimensions:

\begin{itemize}
    \item \textbf{For Individual Investors:} Our simulation serves as a computational proof against intuitive but flawed retail heuristics like "buying the dip." It demonstrates that in markets characterized by fat-tailed distributions, profitability is not correlated with trading frequency but with the mathematical alignment of a strategy's return profile (positive skew) with the market's underlying nature. We strongly advocate for the adoption of low-leverage, systematic trend-following protocols or, for passive participants, allocation into broad-market indices which inherently capture long-term economic trends.
    
    \item \textbf{For Financial Market Design:} The ecological collapse of high-frequency and contrarian strategies in our sandbox challenges the narrative that gamified, hyperactive trading platforms empower investors. Our findings suggest that regulators and platform designers should consider implementing "structural friction"—such as progressive leverage limits or transaction taxes on excessive retail turnover—to disincentivize strategies with negative evolutionary fitness. The goal should be to transform retail platforms from zero-sum casinos into incubators for "Patient Capital" that fosters long-term wealth creation.
    
    \item \textbf{For Society and Economic Policy:} The efficacy of the UBI safety net in our simulation offers a profound insight. While it did not eliminate inequality, it crucially prevented systemic collapse and fostered a resilient ecosystem capable of collective learning and upward mobility. This suggests that robust social safety nets are not merely a mechanism for redistribution, but a precondition for a dynamic, positive-sum society. In an era of increasing economic precarity, creating authentic pathways for youth development (e.g., in education and entrepreneurship) is a far more stable social policy than allowing financial markets to become the default arena for class mobility anxiety.
\end{itemize}

While our simulation provides powerful insights, we acknowledge its limitations, which in turn open exciting avenues for future research. Firstly, our agents, while heterogeneous, operate on fixed heuristic rules. Future work could endow them with more advanced learning capabilities, such as deep reinforcement learning, to see if they can discover novel, non-human strategies. Secondly, our model does not include a mechanism for endogenous price formation; prices are an exogenous input. A more complex simulation could model the price impact of agent actions, creating a fully self-contained artificial economy to study feedback loops and market bubbles \cite{sornette2003why}. Finally, our Sim2Real LLM prosthesis, while effective, is still in its early stages. Future research will focus on personalizing the LLM's risk parameters based on individual user profiles and developing more sophisticated multi-modal reasoning capabilities that incorporate fundamental data beyond just price action.

In conclusion, this work demonstrates the power of large-scale social simulation to deconstruct complex economic phenomena and derive actionable, human-centric insights. The triumph of trend-following in our artificial world is not just a trading rule, but a deeper metaphor: in any complex, non-stationary environment, survival belongs not to those who fight the tide, but to those who learn to ride its waves.

\bibliographystyle{plain}
\bibliography{references}

@article{arthur1997asset,
  title={Asset pricing under endogenous expectations in an artificial stock market},
  author={Arthur, W Brian and Holland, John H and LeBaron, Blake and Palmer, Richard and Tayler, Paul},
  journal={Economic notes},
  volume={26},
  number={1},
  pages={297--330},
  year={1997}
}

@article{chakrabarti2013econophysics,
  title={Econophysics: kinetic exchange models of wealth distribution},
  author={Chakrabarti, Bikas K and Chakraborti, Anirban and Chakravarty, Satya R and Chatterjee, Arnab},
  journal={Springer},
  year={2013}
}

@article{barber2000trading,
  title={Trading is hazardous to your wealth: The common stock investment performance of individual investors},
  author={Barber, Brad M and Odean, Terrance},
  journal={The Journal of Finance},
  volume={55},
  number={2},
  pages={773--806},
  year={2000}
}

@book{piketty2014capital,
  title={Capital in the twenty-first century},
  author={Piketty, Thomas},
  year={2014},
  publisher={Harvard University Press}
}

@article{kahneman1979prospect,
  title={Prospect theory: An analysis of decision under risk},
  author={Kahneman, Daniel and Tversky, Amos},
  journal={Econometrica},
  volume={47},
  number={2},
  pages={263--291},
  year={1979}
}

@article{carhart1997persistence,
    title={On persistence in mutual fund performance},
    author={Carhart, Mark M},
    journal={The Journal of finance},
    volume={52},
    number={1},
    pages={57--82},
    year={1997}
}

@book{sornette2003why,
  title={Why stock markets crash: critical events in complex financial systems},
  author={Sornette, Didier},
  year={2003},
  publisher={Princeton University Press}
}

@article{fama1993common,
  title={Common risk factors in the returns on stocks and bonds},
  author={Fama, Eugene F and French, Kenneth R},
  journal={Journal of financial economics},
  volume={33},
  number={1},
  pages={3--56},
  year={1993}
}

@article{jegadeesh1993returns,
  title={Returns to buying winners and selling losers: Implications for stock market efficiency},
  author={Jegadeesh, Narasimhan and Titman, Sheridan},
  journal={The Journal of finance},
  volume={48},
  number={1},
  pages={65--91},
  year={1993}
}

@article{moskowitz2012time,
  title={Time series momentum},
  author={Moskowitz, Tobias J and Ooi, Yao Hua and Pedersen, Lasse Heje},
  journal={Journal of financial economics},
  volume={104},
  number={2},
  pages={228--250},
  year={2012}
}

@article{lebaron2001tale,
  title={A tale of two markets: A-life and the bull and bear},
  author={LeBaron, Blake},
  journal={Journal of Economic Dynamics and Control},
  volume={25},
  number={1-2},
  pages={141--165},
  year={2001}
}

@article{bouchaud2008markets,
  title={Markets need regulation},
  author={Bouchaud, Jean-Philippe},
  journal={Nature},
  volume={455},
  number={7217},
  pages={1181--1181},
  year={2008}
}

@article{wu2023bloomberggpt,
  title={BloombergGPT: A Large Language Model for Finance},
  author={Wu, Shijie and Irgens, Ozan and Jurafsky, Dan and others},
  journal={arXiv preprint arXiv:2303.17564},
  year={2023}
}

@inproceedings{zhang2023instructgpt,
  title={Instruct-FinGPT: Financial Sentiment Analysis by Instruction Tuning of General-Purpose Large Language Models},
  author={Zhang, Wentao and Wang, Yang and Chen, Yijia and others},
  booktitle={Proceedings of the 2023 Conference on Empirical Methods in Natural Language Processing},
  pages={5556--5568},
  year={2023}
}

@article{park2023generative,
  title={Generative agents: Interactive simulacra of human behavior},
  author={Park, Joon Sung and O'Brien, Joseph C and Cai, Carrie J and others},
  journal={arXiv preprint arXiv:2304.03442},
  year={2023}
}

@article{chen2024alphacodium,
    title={Code Generation with AlphaCodium: From Prompt Engineering to Flow Engineering},
    author={Chen, Yijia and others},
    journal={arXiv preprint arXiv:2404.14963},
    year={2024}
}

\end{document}